\documentclass[12pt]{article}
\usepackage{graphicx}
\usepackage{color}
\newcommand{\beq}{\begin{equation}}
\newcommand{\eeq}{\end{equation}}
\newcommand{\bea}{\begin{eqnarray}}
\newcommand{\eea}{\end{eqnarray}}

\newcommand{\gsim}{\lower.7ex\hbox{$
\;\stackrel{\textstyle>}{\sim}\;$}}
\newcommand{\lsim}{\lower.7ex\hbox{$
\;\stackrel{\textstyle<}{\sim}\;$}}

\newcommand{\eod}{\end{document}}

\definecolor{verm}{rgb}{0.8,0.1,0.0}

\begin{document}
\thispagestyle{empty}
\vspace*{-22mm}

\begin{flushright}

UND-HEP-13-BIG\hspace*{.08em}03\\

\end{flushright}

\vspace*{1.3mm}

\begin{center}
{\Large {\bf 3-\&4-Body Final States in $B$, $D$ \& $\tau$ Decays about 
Features of New Dynamics  
with CPT Invariance  
\vspace*{4mm}
\\ 
{\em or} 
\vspace*{4mm} \\  "Achaeans outside Troy" ￼}}

\vspace*{19mm}

{\bf I.I.~Bigi$^{a,b}$} \\
\vspace{7mm}
$^a$ {\sl  Centro 
Brasileiro de Pesquisas F\'\i sicas, Rua Xavier Sigaud 150, 22290-180 \\ 
Rio de Janeiro, RJ, Brazil}\\
$^b$ {\sl Department of Physics, University of Notre Dame du Lac}\\
{\sl Notre Dame, IN 46556, USA}

\vspace*{-.8mm}

{\sl email address: ibigi@nd.edu}

\vspace*{10mm}

{\bf Abstract}\vspace*{-1.5mm}\\
\end{center}
The `landscape' of fundamental dynamics has changed even for the `known' matter. The Standard Model has produced at least the leading source of CP violation in $B$ decays; the data have not shown CP asymmetries in $D$ transitions. It needs more data and better technologies to understand the underlying forces. 
Probing three- and four-body final states in $B$ \& $D$ \& $\tau$ decays with better accuracy is crucial about the existence and the features of New Dynamics. Theoretical tools produced about MEP will show 
even more about HEP in the future. We have to work on the {\em correlations} between different final states on several CKM levels and the connection between known matter and Dark Matter in indirect ways. CPT invariance is usable in $D$ and $\tau$ decays.

\vspace*{5mm}

\begin{center}

Talk given at 

Conference "Flavor Physics and CP Violation 2013"

Buzioz, Rio de Janeiro, Brazil, May 19 - 24, 2013

\end{center}

\vspace{3mm}

\hrule

\tableofcontents
\vspace{5mm}

\hrule\vspace{5mm}

\section{Prologue -- Landscapes of CP Asymmetries}

Reminding Gary Larson’s `Far Side' cartoon when comboys were surrounded by indians. 
The defenders of the Standard Model (SM) tell us about signs of New Dynamics (ND): SM can do it – and who cares of the `footstool': neutrino oscillations, huge asymmetry in matter vs. anti-matter, `Dark' whatever it is?  

SM gives 
\begin{itemize}
\item 
at least the {\em leading} source of measured CP violation (CPV) in $B$ decays;
\item 
{\em no} CPV in $\tau$ decays beyond measured $K^0 - \bar K^0$ oscillations;  
\item 
{\em small} CPV in {\em singly} Cabibbo suppressed (SCS) in $D_{(s)}$ decays; 
\item 
close to {\em zero} in {\em doubly} Cabibbo suppressed (DCS) charm ones. 
\end{itemize}
Penguin {\em quark} diagrams deal with {\em inclusive} decays. 
The concept of `duality' connect inclusive final states (FS) measured with hadrons and those with 
quarks (\& gluons) that can be calculated.  
There are real challenges, namely to understand the underlying dynamics that give the measured rates.  

The main point is: we have to probe FS about the existence and features of 
ND with both the best theoretical and experimental tools. One should remember that CP asymmetries 
need only one SM amplitude and one ND amplitudes -- i.e., it gives much higher reach about ND; of course interference has to happen. 

We enter a new era: 
\begin{itemize}
\item
The goal is to go from `accuracy' to `precision'. 

\item
CPT invariance can be {\em used} for CP asymmetries in $D$ \& $\tau$ decays or gives the direction for $B$ ones. 
\end{itemize} 
The outline: Sect.\ref{34body} about 3-\&4-body FS of CP asymmetries in $D$, $\tau$ \& $B$ decays; 
Sect.\ref{CPT} about the impact of CPT invariance; Sect.\ref{PARA} about parameterization of CKM matrix 
through higher order; Sect.\ref{TOOLS} about theoretical tools for treating FS interacting (FSI); 
Sect.\ref{BDBS} gives comments about $B_d \to K \pi$ vs. $B_s \to K\pi$ and Sect.\ref{LESSONS} about some lessons learnt in the 
FPCP2013 conference; finally  resume in Sect.\ref{SUM} and Epilogue in Sect.\ref{THOUGHS}. 

\section{3-\&4-Body FS about CPV \& Impact of `Penguins' }
\label{34body}

Probing FS with two hadrons (including narrow resonances) is not trivial to measure CPV; on the other hand 
one gets `just' numbers. However 3-body FS are described by 2-dimensional plots. For 4-body FS one has even more dimensional landscape. There is a price: it needs much 
more work for experimenters to produce such data with more accuracy, and for theorists to understand the information given by the data. 
Yet there is a prize -- namely finding existence of ND and its (or their) features. 

There are several subtle points:
\begin{itemize}
\item
Penguin diagrams were introducted for $K$ decays to understand the forces needed for 
$\Delta I = 1/2 \gg \Delta I = 3/2$ amplitudes and later about the 
ratio $\epsilon ^{\prime}/ \epsilon $. 
They are given by local operators due to $M_K < m_c \ll m_t$; their impacts are greatly enhanced for 
two pseudo-scalars FS. There is hardly any difference between inclusive and exclusive transitions. 

\item 
The landscapes for $B$ transitions are more complex. Quark diagrams including 
penguin ones lead to local operators and can deal with inclusive rates based on quark-hadron duality. It needs work, but it can be done and did it in many cases for {\em inclusive} CPV with `hard' FSI \cite{JARLSBOOK,KOLYA,CPBOOK}.  

Allow me to show simple example, namely $B^- = [b\bar u] \to d \bar u u \bar u$ in the theorists' 
world; in the real world of hadrons they show up  with measured FS separately of 
$2 \pi$, $4 \pi$ etc. and 
$3 \pi$, $5 \pi$ etc. due to $G$ parity.

One can calculate exclusive FS including penguin {\em diagrams}. They give 
the directions of `soft' FSI with {\em hadrons} at best in a {\em semi-quantitative} way; 
i.e., one cannot deal with local operators. Often such contributions are called `effective penguins'; that are fine for leading amplitudes. 

When we want to probe for {\em non}-leading sources, we have to use other theoretical technologies, 
namely to think about correlations with other FS based on global symmetries like chiral, isospin, $SU(3)_{fl}$ etc. Their violations in exclusive FS are sizable larger than inclusive ones. {\em Exclusive} rates depend on 
$M_{\pi} \ll M_K$. {\em Inclusive} ones depend on 
$m_u, m_d \ll m_s$ -- but more importantly on $m_s \ll \bar \Lambda$ which control the impact of QCD 
less flavour dependant.
 
\item
The situations are even more complex for $D$ decays, while on the other hand easier. 

If one has $c \to u$ penguin diagram with internal beauty quarks, it leads to local operator, and one can calculate it -- but it gives only insignificant impact. However internal $d$ and $s$ with $m_d, m_s \ll m_c$ lines are mostly given by  non-perturbative QCD -- in particular about exclusive FS.  

However CPT invariances are `usable' in $D$ and $\tau$ decays, namely about correlations between different FS like $D^+ \to 3 \pi$ vs. $D^+ \to \pi K \bar K$ or 
$\tau ^- \to \nu \bar K^- \pi^0$ vs. 
$\tau ^- \to \nu \bar K^0 \pi^-$ (or $\tau ^- \to \nu K_S\pi^-$). 

\item
Long time before QCD was found as `the' theory about strong forces, one gave 
predictions based on global symmetries like I-, U- and V-spin as parts of $SU(3)_{fl}$. 
It was known that I-spin violation are much smaller than U- and V-spin ones and also somewhat 
smaller than total $SU(3)_{fl}$. Once the technologies of QCD were applied in many situations, 
it was clear that violation of U- and V-spin symmetries are usually larger in exclusive decays than inclusive ones.

\item
Measured rates depend on the areas of Dalitz plots and the production of the decaying state $P$. However the `local' 
ratios of Dalitz plots of $P$ vs. $\bar P$ do not depend on the production of $P$ vs. 
$\bar P$ (in principle). 
\item 
Measuring `local' CP asymmetries needs much more data than `averaged' one, but gives much more information about the underlying dynamics in time. 
\item 
Surprising sign of direct CPV in $\tau$ decays was given:
\bea
A_{\rm CP}(\tau ^+ \to \bar \nu K_S\pi^+)|_{\rm SM} &=& +(0.36 \pm 0.01) \% \; \; \; \cite{BSTAU} 
\label{TAU1}
\\
A_{\rm CP}(\tau ^+ \to \bar \nu K_S\pi^+ [+\pi^{0\; \prime} {\rm s}])|_{\rm BaBar2012} &=& -(0.36 \pm 0.23 \pm 0.11) \%
\; \; \cite{BABARTAU}
\label{TAU2}
\eea

\item 
Since we have no infinite data (and no infinite time), we have to think and discuss which ways 
are the best depending on the features of ND. 
\item
Collaboration of Hadronic Dynamics/MEP and HEP physicists is very important. 

\end{itemize}
Indirect and direct CPV has been established in 2-body FS of $B_d$; we need more precision and 
probe 3- \& 4-body FS with accuracy.

SM expects small {\em indirect} CPV about $B_s$ transitions. It was said a decade ago 
(look at the history given in Ref.\cite{CPBOOK}) even 
ND cannot produce large CPV; one needs more accuracy there. It is a very good achievement by LHCb collab. 
to establish the first CP asymmetry in $B_s$ decays: 
\beq
A_{CP} (B_s \to K^-\pi^+) = 0.27 \pm 0.04 \pm 0.01 \; \; \; \cite{LHCb1}
\label{BSCPV}
\eeq
Yet it is not surprising to find large {\em direct} CPV from the theoretical side in a qualitative way; 
yet it is surprising to me that it is so large; I will came back to it in Sect.\ref{BDBS}. 

We need precision and probe 3-\& 4-body FS with accuracy -- including the 
{\em topologies} of the asymmetries. 
Furthermore you have to measure {\em correlations} between different FS and understand of which reasons they are based.  

There is no evidence for indirect or direct CPV in 2-body FS in $D$ decays. The data are very consistent with SM predictions which 
give small and less CP asymmetries. ND cannot give large CPV, however sizable ones.  
We need precision there and probe {\em 3-\& 4-body} FS. 

There is evidence for direct CPV in $\tau ^+ \to \bar \nu K_S\pi^+$, see Eqs.(\ref{TAU1},\ref{TAU2}). 
We need precision there and probe hadronic {\em 2- \& 3-body} FS. 

We need accuracy on {\em different} CKM levels and {\em correlations}.

\section{Impact of CPT Invariance}
\label{CPT}

CPT symmetry gives equalities for the masses and widths of 
particles $P$ vs. anti-particles $\bar P$. However that invariance tell us much more about 
the underlying dynamics, namely equalities of different classes of FS due to 
`mixing'/`re-scattering' \footnote{Mostly the words `mixing' and `oscillation' are seen as equivalent, however they are not \cite{CPBOOK}.} in the amplitudes \cite{JARLSBOOK,CPBOOK}: 
\bea
T(P \to a) &=& {\rm exp}(i\delta _a)\left[ T_a + i\sum _{aj \neq a} T_{aj} 
T^{\rm resc}_{aj,a} \right]
\\
T(\bar P \to \bar a) &=& {\rm exp}(i\delta _a)
\left[ T^*_a + i\sum _{aj \neq a} T^*_{aj} 
T^{\rm resc}_{aj,a} \right]
\eea 
Direct CPV is measured with 
\beq
\Delta \Gamma (a) = |T(\bar P \to \bar a)|^2 - |T(P \to a)|^2 = 
4 \sum _{aj \neq a}  T^{\rm resc}_{aj,a}  {\rm  Im} T_a^*T_{aj}
\eeq
CPV has to vanish upon summing over all mixed states $a$ 
due to CPT invariance, since $T^{\rm resc}_{aj,a}$ is symmetric and Im$T_a^*T_{aj}$ 
anti-symmetric:
$\sum_a \Delta \Gamma (a) = 4 \sum_a \sum _{aj \neq a}T^{\rm resc}_{aj,a} 
 {\rm Im} T_a^*T_{aj} = 0$. 
We do not know how to calculate strong FSI: $\Delta \Gamma (a)$ cannot predict direct CPV 
{\em quantitatively} even if only SM gives weak phases. 

CPT symmetry gives relations between CP asymmetries in different channels. Finding CP 
asymmetry in one channel one infers which channel(s) have to compensate asymmetries based on CPT invariance. 
Finally analyzing those decays teach us important lessons about the {\em inner} working of QCD. 
CPT invariance in $D$ and $\tau$ decays is `practical', since a `few' channels can be combined. 

Landscapes are different between $D$ \& $\tau$ decays on one side and $B$ ones on the other side. 
Furthermore one has to deal with different experimental and theoretical challenges: 
\begin{itemize}
\item 
To find non-leading sources for ND one has to deal with large `background' for SM about CP asymmetries 
in $B$ decays; 
the impacts of penguin diagrams are subtle and CPT symmetry gives only `directions', but not more: 
`price' vs. `prize'.  
\item 
SM gives small `background' about CPV in SCS and near zero in DCS $D$ decays; 
the impact of penguin 
diagrams are subtle, but different reasons: `prize' vs. `price'. 
\item
SM gives near zero `background' in $\tau$ transition, and the correlations with the forces producing neutrino oscillations; 
on the other hand there could be correlations with CP asymmetries in $D$ decays: `prize' vs. `price'.

\end{itemize}
It was first suggested to use penguins diagrams about FSI \cite{BSS}. However the situations are more subtle and complex as discussed in Ref.\cite{JARLSBOOK}.

\section{Parameterization of CKM Matrix through ${\cal O}(\lambda ^6)$}
\label{PARA}

PDG and HFAG show also the `exact' CKM matrix with three families of quarks. 
However experimenters and theorists do not use exact CKM matrix as you can see in their 
papers and talks. The pattern can be much more obvious in parameterization, tell us when  
we need more data where and the existence of ND and its features. Now we need precision. 

In Wolfenstein parameterization one gets {\em six} triangles that are combined into three classes with four parameters $\lambda$, $A$, $\bar \eta$ and $\bar \rho$ with 
$\lambda \simeq 0.223$. 
Those are probed and measured in $K$, $B$, $B_s$ and $D$ transitions: $A \sim 1$, but 
the two ones are {\em not} of ${\cal O}(1)$: 
$\bar \eta \simeq 0.34$ and $\bar \rho \simeq 0.13$. It is assumed -- usually without mentioning -- that one applies them with no expansion of $\bar \eta$ and $\bar \rho$. 
Obviously it is a `smart' parameterization with a clear hierarchy.  

Now we need a parameterization of the CKM matrix with more precision for non-leading sources 
in $B$ decays and very small CP asymmetries in $D$ decays with little `background' from 
SM. 
Several `technologies' was given like 
in Ref.\cite{AHN} with $\lambda$ as before, but $f\sim 0.75$, $\bar h\sim 1.35$ and 
$\delta _{\rm QM} \sim 90^o$. Now we get somewhat different six classes, and  it is more 
subtle for CP violation:   
\begin{eqnarray} 
\left(\footnotesize
\begin{array}{ccc}
 1 - \frac{\lambda ^2}{2} - \frac{\lambda ^4}{8} - \frac{\lambda ^6}{16}, & \lambda , & 
 \bar h\lambda ^4 e^{-i\delta_{\rm QM}} , \\
 &&\\
 - \lambda + \frac{\lambda ^5}{2} f^2,  & 
 1 - \frac{\lambda ^2}{2}- \frac{\lambda ^4}{8}(1+ 4f^2) 
 -f \bar h \lambda^5e^{i\delta_{\rm QM}}  &
   f \lambda ^2 +  \bar h\lambda ^3 e^{-i\delta_{\rm QM}}   \\
    & +\frac{\lambda^6}{16}(4f^2 - 4 \bar h^2 -1  ) ,& -  \frac{\lambda ^5}{2} \bar h e^{-i\delta_{\rm QM}}, \\
    &&\\
 f \lambda ^3 ,&  
 -f \lambda ^2 -  \bar h\lambda ^3 e^{i\delta_{\rm QM}}  & 
 1 - \frac{\lambda ^4}{2} f^2 -f \bar h\lambda ^5 e^{-i\delta_{\rm QM}}  \\
 & +  \frac{\lambda ^4}{2} f + \frac{\lambda ^6}{8} f  ,
  &  -  \frac{\lambda ^6}{2}\bar h^2  \\
\end{array}
\right)
+ {\cal O}(\lambda ^7)
\end{eqnarray}
\bea
{\rm Class\; I.1:}&&V_{ud}V^*_{us} \; \; \;  [{\cal O}(\lambda )] + V_{cd}V^*_{cs} \;  \; \;  [{\cal O}(\lambda )] + 
 V_{td}V^*_{ts} \; \; \; [{\cal O}(\lambda ^{5\& 6} )] = 0   \\ 
{\rm Class\; I.2:}&& V^*_{ud}V_{cd} \; \; \;  [{\cal O}(\lambda )] + V^*_{us}V_{cs} \; \; \;  [{\cal O}(\lambda )] + 
V^*_{ub}V^*_{cb} \; \; \; [{\cal O}(\lambda ^{6 \& 7} )] = 0    \\
{\rm Class\; II.1:}&& V_{us}V^*_{ub} \; \; \;  [{\cal O}(\lambda ^5)] + V_{cs}V^*_{cb} \;  \; \;  [{\cal O}(\lambda ^{2 \& 3} )] + 
V_{ts}V^*_{tb} \; \; \; [{\cal O}(\lambda ^2  )] = 0   \\ 
{\rm Class\; II.2:}&& V^*_{cd}V_{td} \; \; \;  [{\cal O}(\lambda ^4 )] + V^*_{cs}V_{ts} \; \; \;  [{\cal O}(\lambda ^{2\& 3})] + 
V^*_{cb}V^*_{tb} \; \; \; [{\cal O}(\lambda ^{2 \& 3} )] = 0  \\
{\rm Class\; III.1:}&& V_{ud}V^*_{ub} \; \; \;  [{\cal O}(\lambda ^4)] + V_{cd}V^*_{cb} \;  \; \;  [{\cal O}(\lambda ^{3\& 4} )] + 
V_{td}V^*_{tb} \; \; \; [{\cal O}(\lambda ^3  )] = 0   \\ 
{\rm Class\; III.2:}&& V^*_{ud}V_{td} \; \; \;  [{\cal O}(\lambda ^3 )] + V^*_{us}V_{ts} \; \; \;  [{\cal O}(\lambda ^{3\& 4})] + 
V^*_{ub}V^*_{tb} \; \; \; [{\cal O}(\lambda ^4 )] = 0
\eea 
One finds the same pattern as from Wolfenstein parametrization, namely `large' CP asymmetries in Class III.1, 
sizable ones in Class II.1 and `small' one in Class I.1. However, the pattern is not so obvious, and it is 
similar in a semi-quantitive way:  
\begin{itemize}
\item 
CP asymmetries in $B_d \to \psi K_S$ and $B^+ \to D_+K^+$ control Class III.1 triangle. 
Due to interference between the two contributions one gets from CKM dynamics:
\bea
S(B_d \to \psi K_S)= \sin2\phi_1 &\simeq& 0.62 - 0.68   \; \; {\rm for} \; \delta _{\rm QM} \simeq 75^o - 90 ^o 
\label{AYAN1}
\\
S(B_d \to \psi K_S)= \sin2\phi_1 &\sim& 0.72   \; \; {\rm for} \; \delta _{\rm QM} \simeq 100^o - 120^o \; ; 
\label{AYAN2}
\eea
i.e., CKM dynamics produce 
$S(B_d \to \psi K_S) \sim 0.72$ as largest value for CP asymmetry with 
$\delta _{\rm QM} \simeq 100^o - 120^o$ to compare with the measured
\beq 
S(B_d \to \psi K_S) \sim 0.676 \pm 0.021 \; .
\eeq
Therefore it seems at first sight that CKM dynamics give very close to `maximal' value possible there, but not close to 100 \%. However the situation is more subtle as mentioned next.  

\item 
We are searching for non-leading source of CP violation in $B$ transitions, in particular in 
$B^0 - \bar B^0$ oscillations. 
ND's impact could `hide' there in "SM predicted" CP asymmetries. 
`Data' given by HFAG, for example, are averaged over values of $|V_{ub}/V_{cb}|$ from $B \to l \nu \pi$ and $B\to l \nu X_c$; actually the `central' value is closer to $|V_{ub}|_{\rm excl}$ rather than the larger 
$|V_{ub}|_{\rm incl}$. It is quite possible that the theoretical uncertainties about extracting of 
$|V_{cb}|$, $|V_{ub}|$ and 
$|V_{ub}/V_{cb}|$ from $B \to l \nu \pi$ vs. $B \to l \nu D^*$ are sizably larger than claimed; 
some details are told about it in Ref.\cite{SCH}.

\item
 The information from the data now and in the future about ND has to be based on 
accuracies and its correlations with different FS in several $B$, $D$ and $K$ transitions and rare 
decays. 

\item 
It gives more deeper insight into flavour dynamics and QCD's impact, but also about inner structures for 
non-perturbative forces. 
\end{itemize}

\section{Theoretical Tools for dealing with FSI}
\label{TOOLS}

The goal is to find its (or theirs) existence and its nature. When the impact of ND has been 
established, one wants to find its features due to interferences between scalar \& pseudoscalar, 
vectors \& axial-vectors etc. etc. 
\begin{itemize}
\item 
It is {\em non}-perturbative QCD that mostly controls FSI.
\item 
One has to probe CPV in $K$, $D$, $B$ and $\tau$ decays. 
\item 
There is experience from Hadronic Dynamics(HD)/MEP.

\end{itemize}

\subsection{`Catholic' Road to ND -- 3-Body FS}
\label{CATH}

For $D/B \to P_1P_2P_3$ or $\tau \to \nu P_1P_2$ decays there is a single path to `heaven', 
namely asymmetries in the Dalitz plots. One can rely on {\em relative} rather than 
{\em absolute} CPV; it is much less 
dependent on production asymmetries. However one needs a lot of statistics -- and 
robust pattern recognition\footnote{You might remember the known history.}. 

\subsubsection{CP Asymmetries in $B^{\pm}$ Decays}

One such procedure have given and simulated about 3-body FS in $B^{\pm}$ decays, namely Refs.\cite{MIRANDA1,MIRANDA2};  
another one can be found in Ref.\cite{Williams}.  

Early data from LHCb have found CPV averaged in $B^{\pm}$ decays to 3-body FS \cite{JUSS}:
\bea
A_{CP}(B^{\pm} \to \pi^{\pm} \pi^+\pi^-) &=& + 0.120 \pm 0.020({\rm stat}) \pm 0.019({\rm syst})
\pm 0.007(J/\psi K^{\pm}) \\
A_{CP}(B^{\pm} \to \pi^{\pm} K^+K^-) &=& - 0.153 \pm 0.046({\rm stat}) \pm 0.019({\rm syst})
\pm 0.007(J/\psi K^{\pm})
\eea
It is interesting -- but not more (yet) -- that these CP asymmetries come with opposite signs. It makes it `easier' to think about the impact of CPT invariance. 
 
Very recent data from LHCb show very sizable averaged CP asymmetries 
with more accuracy, correlations and isospin symmetry\cite{LHCb028}: 
\bea
A_{CP}(B^{\pm} \to K^{\pm} \pi^+\pi^-) &=&  
+0.034 \pm 0.009({\rm stat}) \pm 0.004({\rm syst})
\pm 0.007(J/\psi K^{\pm})  \\
A_{CP}(B^{\pm} \to K^{\pm} K^+K^-) &=&  
- 0.046 \pm 0.009({\rm stat}) \pm 0.005({\rm syst})
\pm 0.007(J/\psi K^{\pm}) \; .
\eea
The data of CKM suppressed $B^+$ decays to charged three-body FS 
\bea
{\rm BR}(B^+ \to K^+\pi^-\pi^+) &=&(5.10 \pm 0.29 ) \cdot 10^{-5}  \\  
{\rm BR}(B^+ \to K^+K^-K^+) &=&(3.37 \pm 0.22 ) \cdot 10^{-5} \\
{\rm BR}(B^+ \to \pi^+\pi^-\pi^+) &=&(1.52 \pm 0.14 ) \cdot 10^{-5}
\\ 
{\rm BR}(B^+ \to \pi^+K^-K^+) &=&(0.52 \pm 0.07 ) \cdot 10^{-5}
\eea
show the impact of penguins/re-scattering diagrams, since the FS with $\Delta S \neq 0$ 
are larger than with $\Delta S = 0$. However one can remember that penguins operators show only 
{\em hard} re-scattering and focus on inclusive decays.   

It is important to measure the averaged CP asymmetries, but also probe the Dalitz plots `locally' 
and probe the correlations with different FS as shown above.

\subsubsection{CP Asymmetries in $D^{\pm}_{(s)}$ Decays}

$D^{\pm}$ has two all charged 3-body FS on the SCS level -- namely 
$D^{\pm} \to \pi^{\pm} \pi^+\pi^-$ and $D^{\pm} \to \pi^{\pm} K^+K^-$ \cite{MIRANDA3} -- 
and also on the DCS one -- $D^{\pm} \to K^{\pm} \pi^+\pi^-$ and $D^{\pm} \to K^{\pm} K^+K^-$. 
$D_s^{\pm}$ has two ones on the SCS level -- $D^{\pm}_s \to K^{\pm} \pi^+\pi^-$ and 
$D^{\pm}_s \to K^{\pm} K^+K^-$ -- 
however only one for DCS level -- $D^{\pm}_s \to K^{\pm}K^{\pm}\pi^{\mp}$.

As stated above, for SCS FS SM gives small `background' for CPV and close to zero about DCS. However data give limits about CPV 
that are somewhat small or happen in rare FS. We have to use good experimental and theoretical technologies to get the information about the underlying dynamics; 
we have to probe FS with broad resonances -- in particular scalar ones like $\sigma$ and $\kappa$ -- and their interferences.

\subsubsection{CP Asymmetries in $\tau^{\pm}$ Decays and Correlations with $D_{(s)}$ Decays}

There may be a sign - may be -- of ND in $\tau$ decays, see Eqs. (\ref{TAU1},\ref{TAU2}) about 
{\em averaged} CP asymmetries. 
It is crucial to probe CP `locally'. Furthermore one has to measure correlations with $D^{\pm}_{(s)}$ decays \cite{NAGOYAIB}.

One should focus on SCS decays with the impact with two hadrons in the FS, namely $\tau ^- \to \nu K^+\pi^0$ and $\tau ^+ \to \nu K_S\pi^+$ and FS with more hadrons.

\subsection{`Protestant' Road to ND -- 4-Body FS}
\label{PROT}

There are several ways to probe CPV in 4-body FS and to differential the impact of SM vs. ND, since the landscapes are more complex. One can compare 
T {\em odd} moments or correlatios in $D$ vs. $\bar D$.  For example one has to measure the angle $\phi$ between the 
planes of $\pi^+-\pi^-$ and $K - \bar K$ and described its dependence 
\cite{CPBOOK, NAGOYAIB}: 
\bea
\frac{d\Gamma}{d\phi} (D \to K \bar K \pi^+\pi^-) &=& \Gamma_1 {\rm cos^2}\phi + \Gamma_2 {\rm sin^2}\phi +\Gamma_3 {\rm cos}\phi {\rm sin}\phi
\\ 
\frac{d\Gamma}{d\phi} (\bar D \to K \bar K \pi^+\pi^-) &=& \bar \Gamma_1 {\rm cos^2}\phi + \bar \Gamma_2 {\rm sin^2}\phi -\bar \Gamma_3 {\rm cos}\phi {\rm sin}\phi
\eea
The partial width for $D[\bar D] \to K \bar K \pi^+\pi^-$ is given by $\Gamma _{1,2} [\bar \Gamma _{1,2}]$: $\Gamma_1 \neq \bar \Gamma_1$ 
and/or  $\Gamma_2 \neq \bar \Gamma_2$ represents direct CPV in the partial width. 

$\Gamma_3$ and $\bar \Gamma_3$ represent T {\em odd} correlations; 
by themselves they do not necessarily indicate CPV, since they can be induced by strong FSI; however \cite{RIOMANI,TODD,CPBOOK}:
\beq 
\Gamma_3 \neq \bar \Gamma_3  \; \; \to \; \; {\rm CPV} 
\eeq   
Integrated rates give $\Gamma_1+\Gamma_2$ vs. $\bar \Gamma_1 + \bar \Gamma_2$; 
integrated {\em forward-backward} asymmetry 
\beq
\langle A\rangle = 
\frac{\Gamma_3 - \bar \Gamma_3}{\pi /2(\Gamma_1+\Gamma_2+\bar \Gamma_1+\bar \Gamma_2)}
\eeq
gives full information about CPV. One could disentangle $\Gamma_1$ vs. $\bar \Gamma_1$ and 
$\Gamma_2$ vs. $\bar \Gamma_2$ by tracking the distribution in $\phi$.

\subsection{Theoretical Tools for treating FSI}
\label{NABIS}

Tools about FSI in 3- and 4-body FS have been produced after the last 10 - 15 years mostly based on dispersion relations. See a list of papers \cite{DISPREL}. There are some points: 
chiral symmetry is a good tool for probing FS with just pions, but not about $D$ and 
$\tau$ decays with kaons. However the connection of CPT and chiral symmetries is subtle.

\section{CP Asymmetry in $B_s \to K^-\pi^+$ vs. $B_d \to K^+\pi^-$}
\label{BDBS}

It is an important achievement that LHCb has found the first CP asymmetry in $B_s$ decays, and it is large: $A_{\rm CP}(B_s \to K^-\pi^+) = 0.27 \pm 0.04 \pm 0.01$. There is an obvious 
reason to compare it with 
$A_{\rm CP}(B_d \to K^+\pi^-) = -0.080 \pm 0.007 \pm 0.003$. The correlation of those 
CP asymmetries come with opposite signs is not surprising. Furthermore it gives 
with an amazing {\em experimental} certainty 
$\Delta _{\rm LHCb} = - 0.02 \pm 0.05 \pm 0.04$ 
\cite{LHCbPAPER13018}. 
There is a statement just before the Eq.(11): "These results allow a stringent test of the validity of the relation between $A_{CP}(B^0 \to K^+\pi^-)$ and $A_{CP}(B^0_s \to K^-\pi^+)$ in the SM given in Ref.\cite{LIPKIN}. However in the 2005 paper Lipkin gave a theoretical uncertainty of 
`... the order of 10 - 20 per cent ...' on p. 6 of arXiv paper. It is not enough to just read the `Abstract' of Lipkin's papers, but more. In this paper the prediction was based on a {\em model} about CPT 
invariance that is stated in the body of that paper. CPT 
symmetry is `practical' about charm and $\tau$ decays, but for beauty decays only about `directions'. 
Furthermore {\em measured} FS in $B$ decays are based on hadrons, not quarks. Theory tells about underlying 
forces for inclusive transitions in a quantitative way including `hard' re-scattering due to the total QCD.   
The challenge for exclusive ones is greater: the impact of strong 
FSI is important for {\em correlations} between members of the same class defined by symmetries, but  
not by the numbers of hadrons; however we cannot calculate 
them \cite{JARLSBOOK,KOLYA,CPBOOK}.  Therefore it has to probe $B$ decays with three- and four-body FS in the future with `local' CP asymmetries  -- in particular about regions where ND can have more impact like from exchanges with charged Higgs etc. 
Furthermore SM gives at least the leading source of beauty CPV.

One more comment: for a long time it was stated that Cabibbo flavoured quark penguin loop diagrams can compete with Cabibbo suppressed 
quark tree diagrams on a similar level -- even more -- and produre interferences leading to direct CP violation on the scale of around 10 \% -- as shown for $B_d \to K^+\pi^-$ as suggested in Ref.\cite{BSS,JARLSBOOK}. However the situation is quite different for $B_s \to K^-\pi^+$, where one have 
to compare Cabibbo favoured quark tree with Cabibbo suppressed quark penguins loop diagrams. As stated 
before, quark diagrams deal with inclusive decays: $B_d \to K X_{S=0}$ 
vs. $B_s \to \bar K X_{S=0}$.  The impact of strong forces due to re-scattering 
is important or even crucial for exclusive hadronic FS. The data tell us that the rates of 
$B_s \to K^- \pi^+$ are smaller than for $B_d \to K^+\pi^-$, but not more than a factor of two or three. 
It seems to me that might be a sign of ND's impact there, unless there is an impact of a resonance. 
Obviously I need more thinking about it based on the concept of `duality' in subtle ways \cite{DUAL}. 

\section{Lessons from FPCP2013 Conference in `Person'}
\label{LESSONS}

Talking in person is much more important than connection by internet about fundamental physics. 
It helps to understand items covered in talks at conferences where one can think about and 
discuss them not only with the speaker, but with other attendees. 

One example: BaBar/Belle data show that $B$ decays have probability with baryon-antibaryon FS with 
$(6.8 \pm 0.6)\%$. 
Based on a parton model a 1981 prediction gave $(5 - 10)\%$ in the range for these {\em inclusive} FS \cite{BIGI81}. It is surprising about these 
data show that known exclusive FS give only 10 \% of these inclusive ones. It tells us that our control of non-perturbative 
dynamics is quite limited. We have to think about the impact of resonances (in particular broad one), threshold enhancement etc. Of course, there is no other candidate about strong forces different from QCD. It shows that we have to think more about impact of ND in hadrons decays. 
It would not be surprising that {\em semi-leptonic} FS with baryon-antibaryon are less complex than those non-leptonic ones. 

Finally the central point is how important meetings, conferences and workshop are even in the internet era.

\section{Summary of Searching for ND in 3-\&4-Body FS}
\label{SUM}

The goal of flavor dynamics is to find the {\em existence} and {\em features} of ND. The SM with 
$SU(3)_{\rm QCD} \times SU(2)_L \times U(1)$ is not complete even beyond thinking about symmetries -- 
namely the structure of our Universe, neutrino oscillations, asymmetry in matter vs. anti-matter. It has a good 
chance to show the correlation between known matter vs. dark matter in heavy flavor transitions. Therefore we have to probe 3-\&4-body FS of $D$, $B$ \& $\tau$ decays with 
{\em correlations}. We need detailed analyses of 
3-\&4-body FS including CPV despite the large start-up work. CPT invariance does not have 
just an academic reason, but also a practical one at least in $D$ and $\tau$ decays.
We should have real collaborations between theorists from HD/MEP \& HEP and experimentalists from HEP. 
It is important whether penguin quark diagrams lead to a local operator or not. We have to remember  
that U- and V-spin violations enter different landscapes in exclusive vs. inclusive decays. 

Most physicists start with minimal version of ND for practical reasons; however the real world does not care about convenience for our powers of calculations. The best example is SUSY: there are several causes for the existence of SUSY in our world -- however those do not give us reasons for minimal version of SUSY or close to it. 

\vspace*{5mm}

{\bf A few words {\em after} this Conference:} A very, very recent paper from LHCb collab. 
is dealing with $B^0_{(s)} \to K_Sh^+h^{\prime  -}$ with $h=\pi,K$ \cite{AFTER}. 
It needs more thinking about the informations that the data give us about CKM suppressed $B^0_{(s)}$ rates including re-scattering -- and about CP asymmetries, signs of ND existence and its features. 
This landscape of three-body FS seems to be more `complex' based on quark diagrams.  

\section{Epilogue: `Achaeans outside Troy'}
\label{THOUGHS}

\begin{figure}[h!]
\begin{center}
\includegraphics[width=12.5cm]{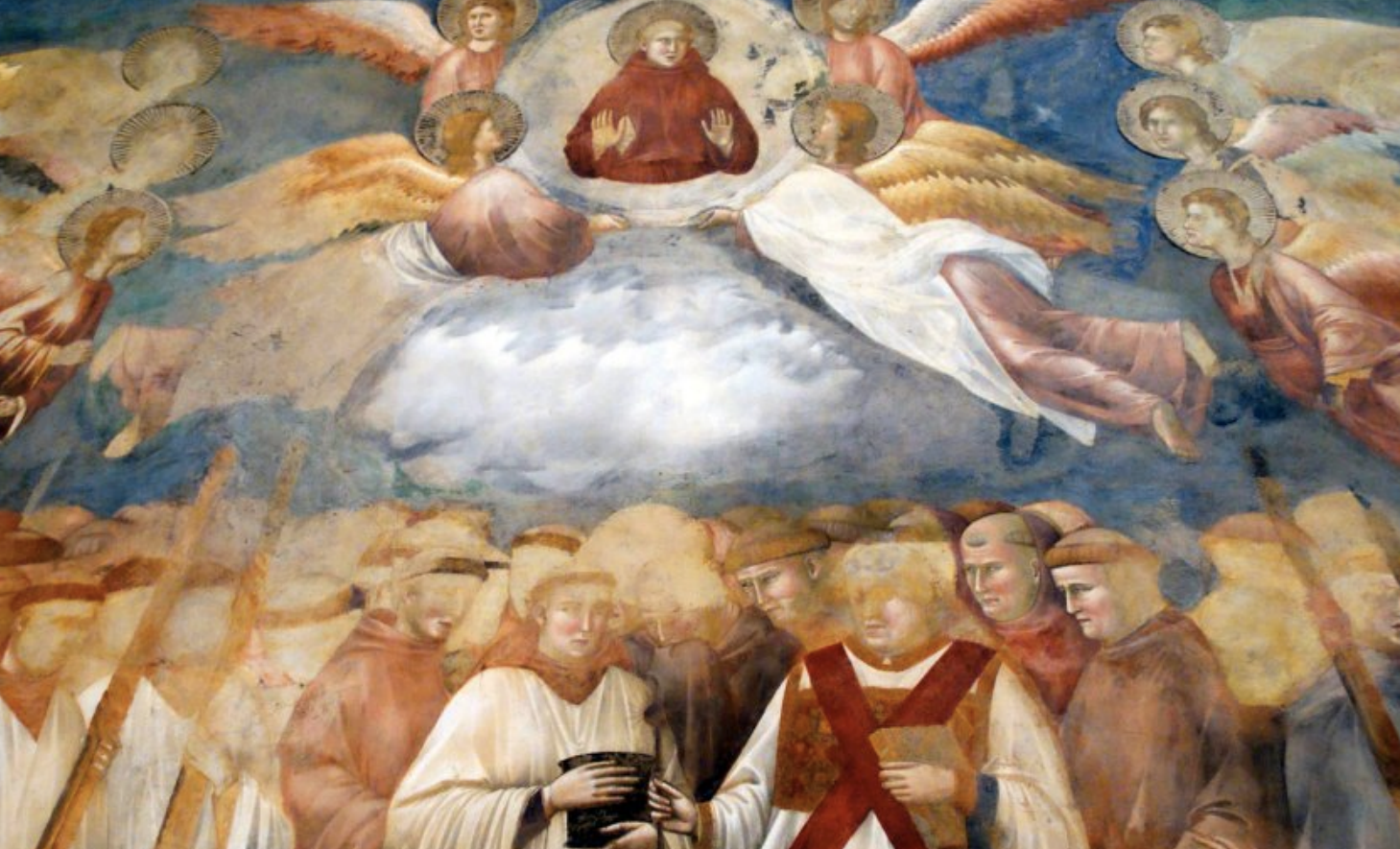}
\end{center}
\caption{Painting in the Basilica San Francesco in Assisi in Italy}
\label{fig:DEV1}
\end{figure}

\begin{figure}[h!]
\begin{center}
\includegraphics[width=12.5cm]{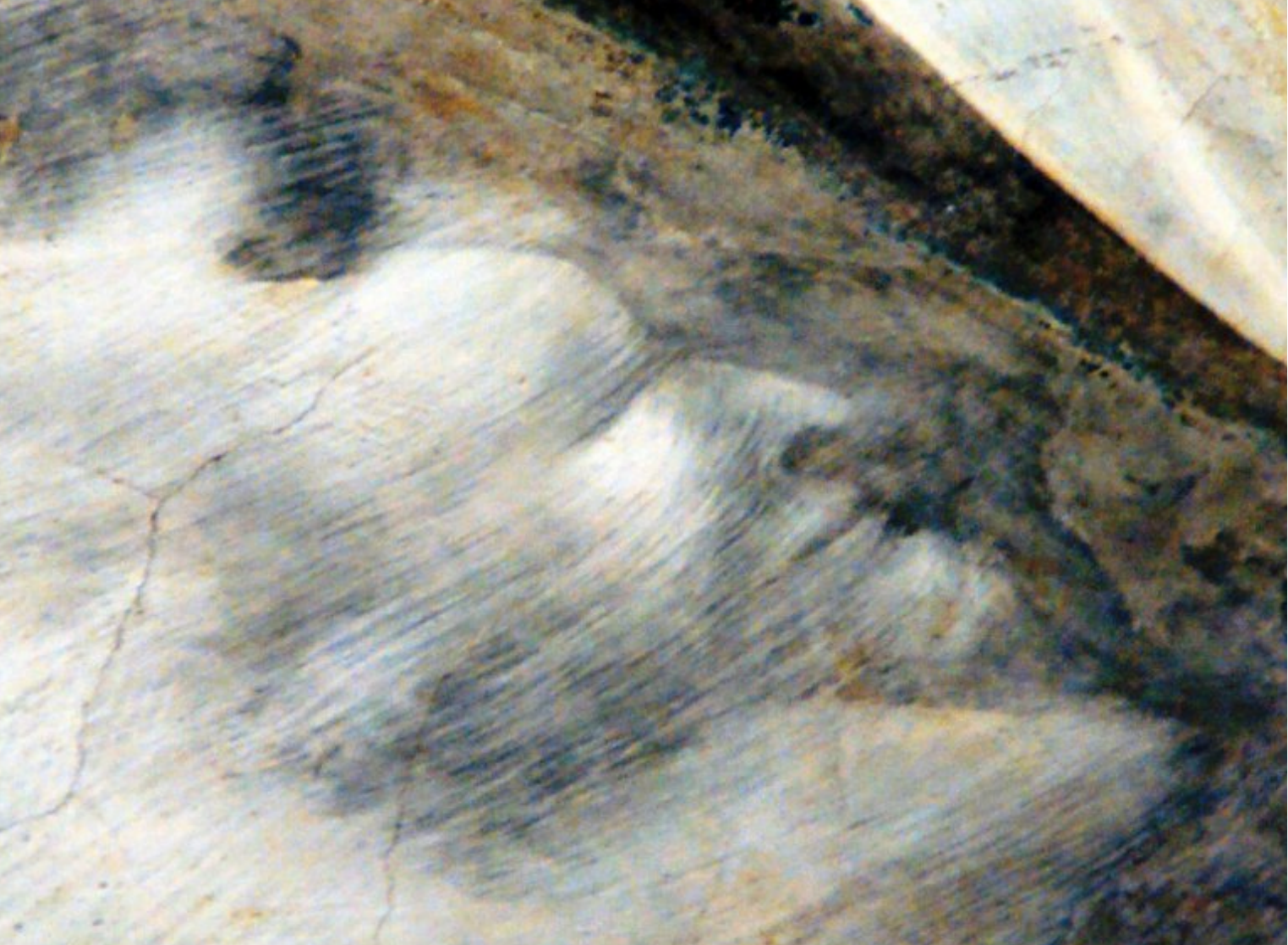}
\end{center}
\caption{Finding the Devil in the Basilica San Francesco in Asssi in Italy}
\label{fig:DEV2}
\end{figure}

\begin{figure}[h!]
\begin{center}
\includegraphics[width=13.5cm]{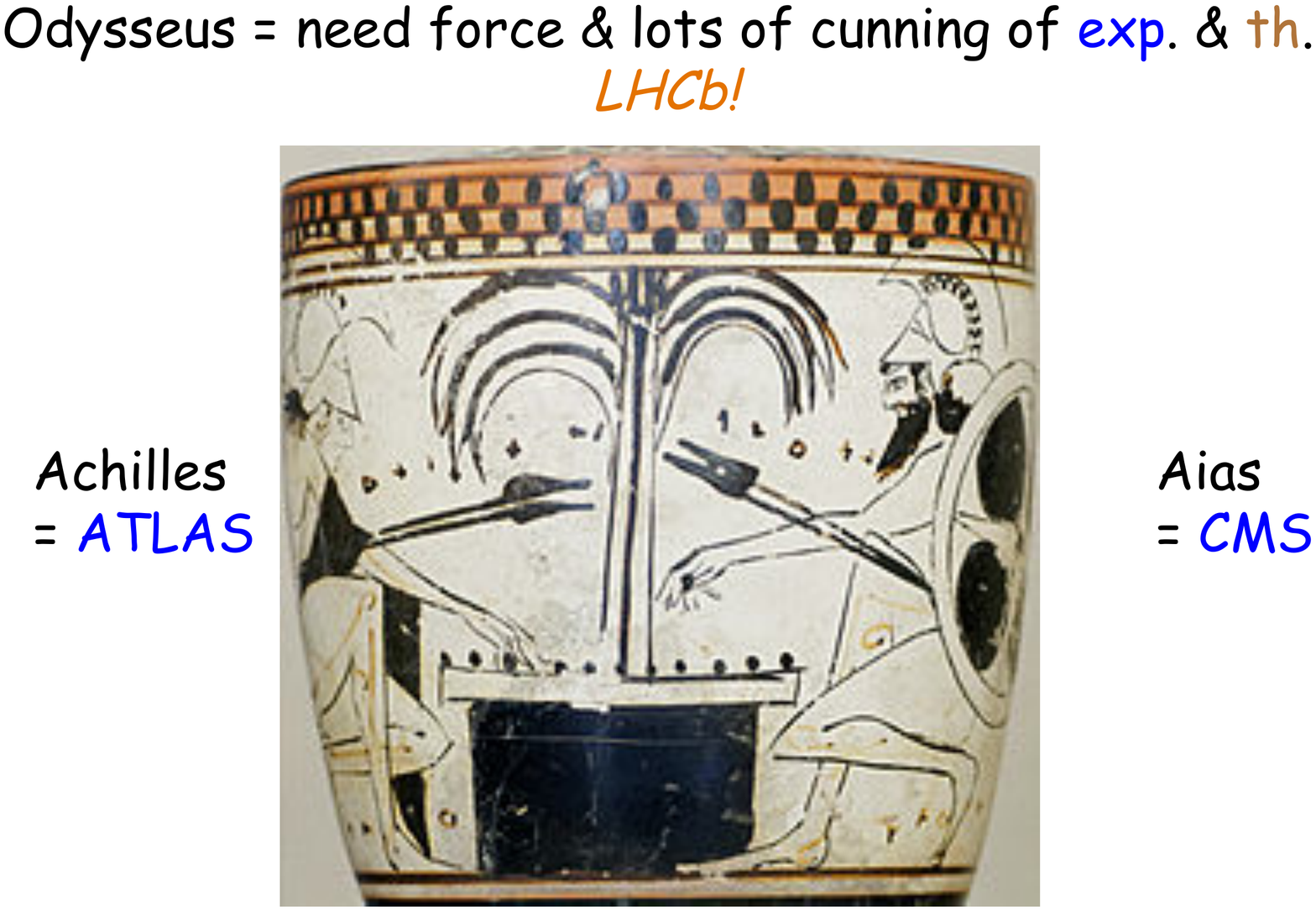}
\end{center}
\caption{Achaeans needed force, but also `cunning' to win -- and Odysseus produced both 
for taking Troy!}
\label{fig:ODY}
\end{figure}

It was said for a very, very long time that one could find the `Devil' at least in one paintings produced 
by Giotto in the Basilica San Francesco in Assisi in Italy in the 14th century, see Fig.\ref{fig:DEV1}. 
A few years ago it was found 
in a subtle locality. Obviously it took many efforts to find `him' even with his horn, beard and strong nose. Now you can see him 
in the cloud \footnote{Have you heard the word `Cloud' more recently in another, but still `sacred' situation?}, see Fig.\ref{fig:DEV2}.
There is a much longer history that painting, namely the Greek history about taking Troy. We hope that the features of ND will be seen and measured in the next ten years with the modern analogy, see the Fig.\ref{fig:ODY}. 

{\bf Acknowledgments:} This work was supported by the NSF under the grant number PHY-0807959 and by CNPq. 
I have really enjoyed the FPCP2013 conference for several reasons led by Leandro Salazar De Paula.
 For a long time I have benefitted very much with the collaboration 
with J. Miranda, I. Bediaga and A. Reis. A. Paul helped me for producing the numbers in 
Eqs.(\ref{AYAN1}),(\ref{AYAN2}).



\end{document}